\documentclass[aps, prl, twocolumn, letterpaper, superscriptaddress]{revtex4-1}

\usepackage{amsmath}
\usepackage{amssymb}
\usepackage{xspace}
\usepackage{color}
\usepackage{graphicx}
\usepackage{hyperref}
\usepackage{ulem}
\usepackage{ifpdf}
\ifpdf
\fi

\newcommand{\eq}[1]{(\ref{#1})}
\newcommand{\Eq}[1]{Eq.~(\ref{#1})}

\newcommand{\Fig}[1]{Fig.~\ref{#1}}

\newcommand{\Ref}[1]{Ref.~\cite{#1}}
\newcommand{\Refs}[1]{Refs.~\cite{#1}}

\newcommand{\eg}{{e.g.,\/}\xspace}
\newcommand{\ie}{{i.e.,\/}\xspace}
\newcommand{\etal}{{\it et~al.\/}\xspace}

\newcommand{\pd}{\partial}
\newcommand{\del}{\nabla}
\newcommand{\mc}[1]{\mathcal{#1}}
\newcommand{\bd}[1]{\boldsymbol{#1}}

\renewcommand{\vec}[1]{{\bd{#1}}}

\newcommand{\favr}[1]{\langle #1 \rangle}

\begin{document}

\title{Theory of the tertiary instability and the Dimits shift from reduced drift-wave models}

\author{Hongxuan Zhu}
\affiliation{Princeton Plasma Physics Laboratory, Princeton, NJ 08543}
\affiliation{Department of Astrophysical Sciences, Princeton University, Princeton,
NJ 08544}

\author{Yao Zhou}
\affiliation{Princeton Plasma Physics Laboratory, Princeton, NJ 08543}

\author{I.~Y. Dodin}
\affiliation{Princeton Plasma Physics Laboratory, Princeton, NJ 08543}
\affiliation{Department of Astrophysical Sciences, Princeton University, Princeton,
NJ 08544}

\begin{abstract}
Tertiary modes in electrostatic drift-wave turbulence are localized near extrema of the zonal velocity $U(x)$ with respect to the radial coordinate $x$. We argue that these modes can be described as quantum harmonic oscillators with complex frequencies, so their spectrum can be readily calculated. The corresponding growth rate $\gamma_{\rm TI}$ is derived within the modified Hasegawa--Wakatani model. We show that $\gamma_{\rm TI}$ equals the primary-instability growth rate plus a term that depends on the local  $U''$; hence, the instability threshold is shifted compared to that in homogeneous turbulence. This provides a generic explanation of the well-known yet elusive Dimits shift, which we find explicitly in the Terry--Horton limit. Linearly unstable tertiary modes either saturate due to the evolution of the zonal density or generate radially propagating structures when the shear $|U'|$ is sufficiently weakened by viscosity. The Dimits regime ends when such structures are generated continuously.
\end{abstract}

\maketitle

Drift-wave (DW) turbulence plays a significant role in fusion plasmas and can develop from various ``primary'' instabilities \cite{Horton99,Jenko00,Rogers00,Hasegawa77,Hasegawa83,Kobayashi12,Leconte19}. However, having their linear growth rate~$\gamma_{\rm PI}$ above zero is not enough to make plasma turbulent, because the ``secondary'' instability can suppress  turbulence by generating zonal flows (ZFs) \cite{Rogers00,Jenko00,Leconte19,Mikkelsen08,Lin98,Diamond01,Biglari90,Diamond05}; hence, the threshold for the onset of turbulence  is modified compared to the linear theory. This constitutes the so-called Dimits shift \cite{Dimits00,St-Onge17,Numata07,Kolesnikov05,Mikkelsen08,Ricci06}, which has been attracting  attention for two decades. The finite value of the Dimits shift is commonly attributed to the ``tertiary'' instability (TI) \cite{Rogers00}, and some theories of the TI have been proposed \cite{Rogers00,Ricci06,Rogers05,St-Onge17,Numata07,Kim02,Rath18}. However, basic understanding and generic description of the TI and the Dimits shift have been elusive.

Here, we propose a simple yet quantitative theory of the TI using the modified Hasegawa--Wakatani equation (mHWE) \cite{Hasegawa83, Numata07} as a base turbulence model. We clarify several misconceptions regarding the TI, and we explicitly derive the Dimits shift in the limit corresponding to the Terry--Horton model \cite{Terry82,St-Onge17}. Our approach is also applicable to  other DW models, such as ion-temperature-gradient (ITG)  turbulence \cite{Rogers00}, as discussed towards the end. Furthermore, we explain TI's role in two types of predator--prey (PP) oscillations, in  determining the characteristic ZF scale in the Dimits regime, and in transition to the turbulent state.

\begin{figure}
\includegraphics[width=1\columnwidth]{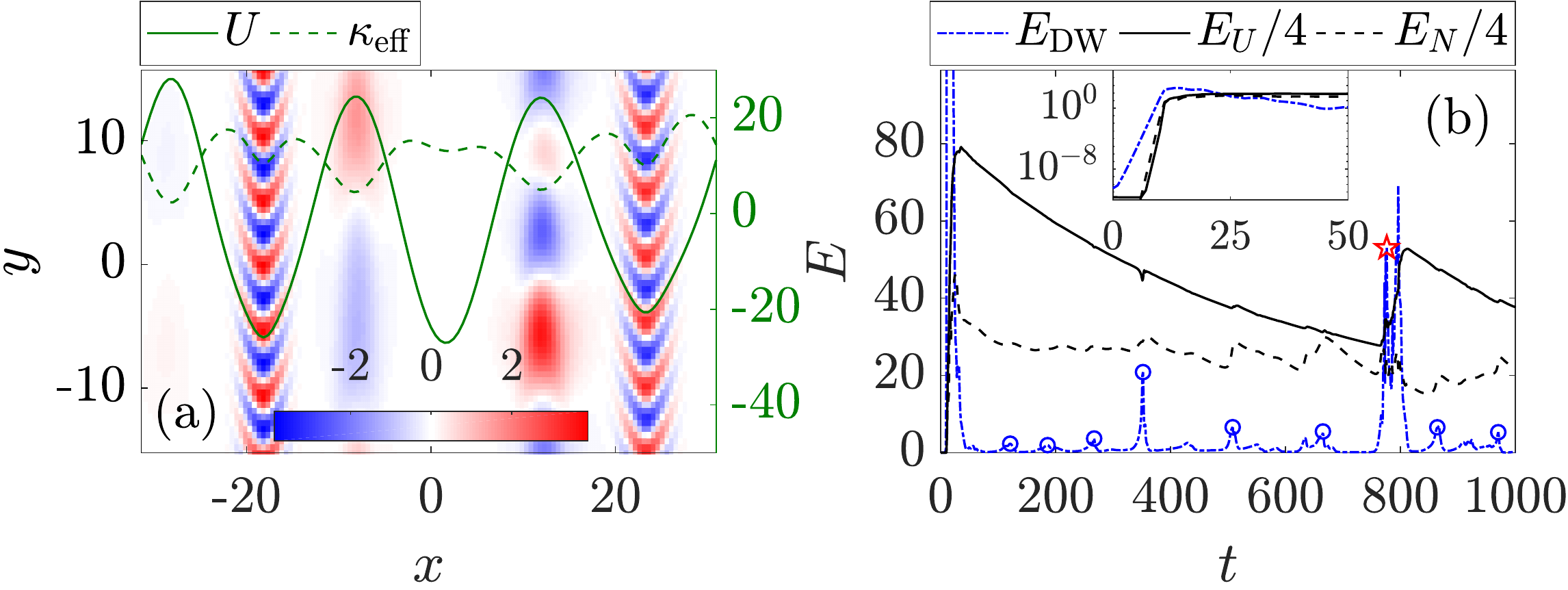}\caption{Results of mHWE simluations using \textsc{Dedalus} \cite{Dedalus} with $\alpha=5$, $\kappa=12$, and $\hat{D}=0.1\nabla^4$. The system is initialized with small random noise. (a) A snapshot at $t=1000$ showing fluctuations $\tilde{w}$ (color), $U$ (solid line), and  $\kappa_{\rm eff}\doteq\kappa-N'$ (dashed line).  (b) Evolution of the DW energy $E_{{\rm DW}}\doteq\int(\tilde{n}^{2}+|\nabla\tilde{\varphi}|^{2}){\rm d}\bd{x}/2$ , and the zonal energy  $E_U=\int U^2{\rm d}\bd{x}/2$ and $E_N=\int N^2{\rm d}\bd{x}/2$. The inset is a zoom-in showing the energies in log scale at $0<t<50$, when the primary and secondary instabilities develop. Type-I PP oscillations are indicated by blue circles, and a type-II PP oscillation is indicated by
 the red star. (See text for more details.)}\label{SnapShot}
\end{figure}

{\it Model equations.---} The mHWE \cite{Hasegawa83, Numata07} is a slab model of two-dimensional electrostatic turbulence with a uniform magnetic field $\bd{B}=B\hat{\bd{z}}$. Turbulence is considered on the plane $(x, y)$, where $x$  is the radial coordinate and $y$ is the poloidal coordinate. The model describes $\varphi$ and $n$, which are fluctuations of the electric potential and density, respectively. Ions are assumed cold while electrons have finite temperature $T_{\rm e}$. The plasma is assumed to have an equilibrium density profile  $n_0(x)$ parameterized by a constant $\kappa\doteq a/L_n$, where $a$ is the system length and $L_n\doteq(-n_0'/n_0)^{-1}$. (We use $\doteq$ to denote definitions and prime to denote $\pd_x$.)  Plasma resistivity produces primary instabilities and is modeled by the  ``adiabaticity parameter'' $\alpha$.  We normalize time by $a/C_{\rm s}$ where $C_{\rm s}\doteq\sqrt{T_{\rm e}/m_{\rm i}}\,$,  length by $\rho_s\doteq C_{\rm s}/\Omega_{\rm i}$, where $\Omega_{\rm i}$ is the ion gyrofrequency, and  $\varphi$ by $T_{\rm e}\rho_{\rm s}/ea$.
Then, the mHWE is written as \cite{Numata07}
\begin{equation}
{\rm d}_tw=\kappa\pd_y\varphi-\hat{D}w,\quad {\rm d}_t n=\alpha(\tilde{\varphi}-\tilde{n})-\kappa\partial_{y}\varphi-\hat{D}n.\label{mHWE}
\end{equation}
Here, ${\rm d}_t\doteq\pd_t+(\hat{\vec{z}}\times\del\varphi)\cdot\del$ and $w\doteq\del^2\varphi-n$ is minus the ion gyrocenter-density perturbation \cite{Krommes00}. Also, $\tilde{\varphi}$ and $\tilde{n}$ are the non-zonal parts of $\varphi$ and $n$; \eg $\tilde{\varphi}\doteq\varphi-\favr{\varphi}$, where $\favr{\dots}$ denotes  average over $y$, or zonal average. The operator $\hat{D}$  models drag and (or) viscosity; its specific form is not essential. (We choose $\hat{D}$ to be   hyperviscosity large compared to that in \Ref{Numata07}, so the related effects manifest in simulations faster and more clearly.)  The zonal average of \Eq{mHWE} reads
\begin{equation}
\pd_{t}U=-\langle\tilde{v}_{x}\tilde{v}_{y}\rangle'-\hat{D}U,\quad\pd_{t}N=-\langle\tilde{v}_{x}\tilde{n}\rangle'-\hat{D}N,\label{Reynolds}
\end{equation}
where $U(x,t)\doteq\langle\varphi\rangle'$ is the ZF velocity, $N(x,t)\doteq\langle n\rangle$ is the  zonal-density perturbation, and $(\tilde{v}_x,\tilde{v}_y)=\hat{\vec{z}}\times\del\tilde{\varphi}$. Below, we develop our TI theory based on \Eq{mHWE}.  An example of mHWE simulations is also illustrated in \Fig{SnapShot} and will be discussed later. 

{\it Tertiary instability.---} Absent ZFs, $w\propto{\rm e}^{{\rm i}(\vec{p}\cdot\vec{x}-\Omega_{\vec{p}}t)}$ is a linear eigenmode whose dispersion relation is similar to that in the Hasegawa--Mima model \cite{Hasegawa77}:
\begin{equation}
\Omega_{\bd{p}}=\frac{\kappa p_y}{\bar{p}^2}-{\rm i}D_{\bd{p}},\quad \bar{p}^{2}\doteq p_x^2+p_y^2+\frac{{\rm i}\alpha+\kappa p_{y}}{{\rm i}\alpha+{\rm i}D_{\bd{p}}+\Omega_{\bd{p}}}. \label{LinearFre}
\end{equation}
Here, $\bd{p}=(p_x,p_y)$, and $D_{\bd{p}}$ is obtained from $\hat{D}$ by replacing $\del$ with $-{\rm i}\bd{p}$. The primary instability develops when $\gamma_{\rm PI}\doteq{\rm Im}\,\Omega_{\bd{p}}>0$, and $\gamma_{\rm PI}$ is maximized at $p_x=0$ and $p_y\sim 1$. However, this instability is  modified once ZFs are generated by the secondary instability. In the presence of ZFs, DWs tend to localize near extrema of  $U$, as seen in \Fig{SnapShot}(a) (also observed in \Refs{Kobayashi12, Kim18a, Kim19}), and their growth rates are also affected. To describe these effects, we assume a zonal state with prescribed $U(x)$ and $N(x)$  and consider a perturbation $\tilde{w}={\rm Re}[\psi(x){\rm e}^{{\rm i}(p_{y}y-\omega t)}]$. Then, the linearized mHWE (\ref{mHWE}) leads to
\begin{equation}
\omega\psi =\hat{H}\psi,\quad\hat{H}(x,\hat{p}_x)=p_{y}[U+(\kappa_{\rm eff}+U'')\hat{\bar{p}}^{-2}]-{\rm i}\hat{D}.\label{eigen_num}
\end{equation}
Here,  $\kappa_{\rm eff}\doteq\kappa-N'$, $\hat{p}_x\doteq -{\rm i} {\rm d}/{\rm d} x$, and
\begin{equation}
\hat{\bar{p}}^{2}\doteq\hat{p}_{x}^{2}+p_{y}^{2}+({\rm i}\alpha+{\rm i}\hat{D}+\omega-p_{y}U)^{-1}({\rm i}\alpha+p_{y}\kappa_{\rm eff}). \label{pbarhat2}
\end{equation}
Note that $\hat{H}$ depends on $\omega$ through $\hat{\bar{p}}$.

\begin{figure}
\includegraphics[width=1\columnwidth]{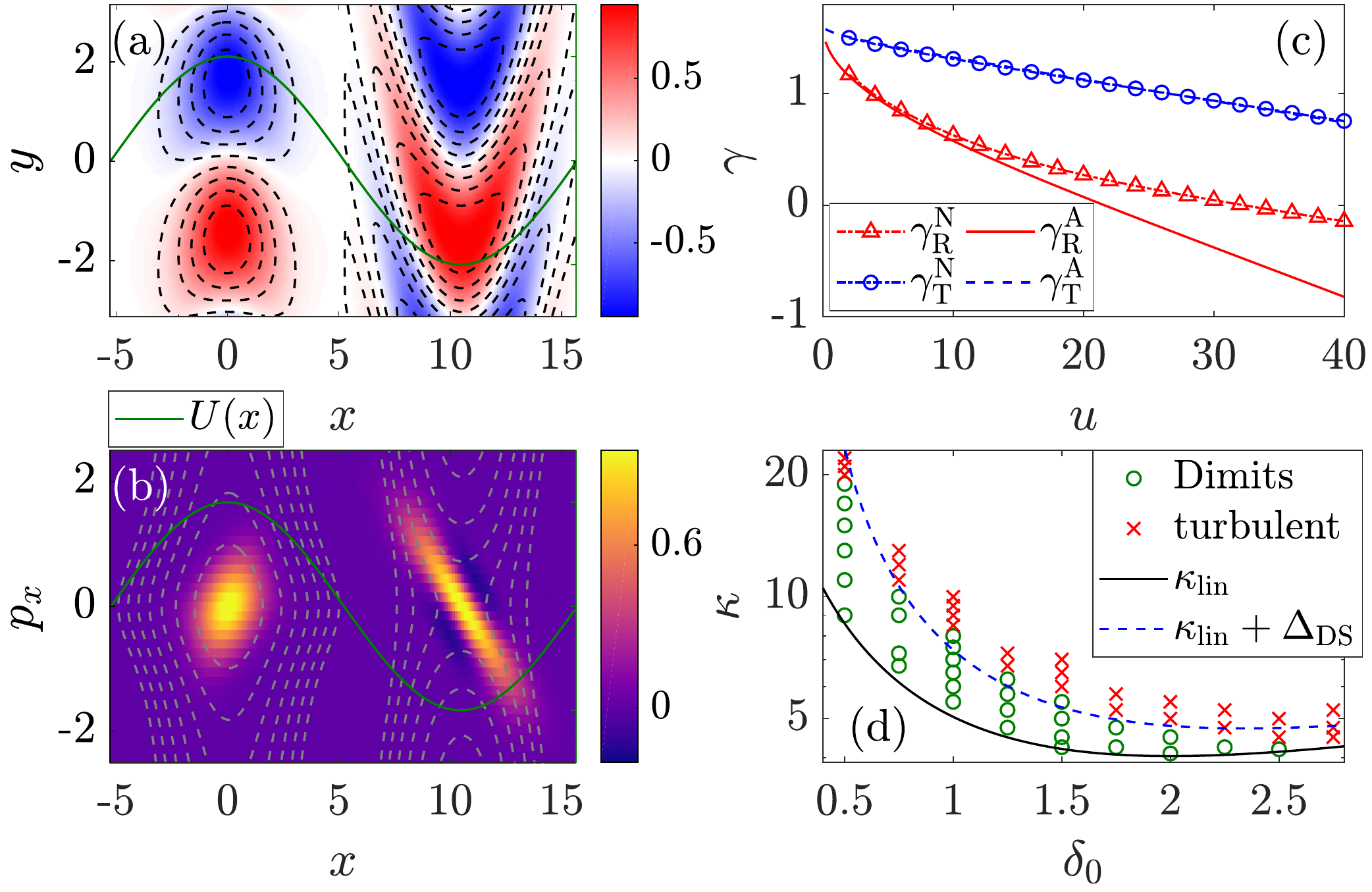}\caption{(a) Trapped and runaway modes (plotted together) for the zonal profile \eq{zonal_zeroN}: numerical solutions of \Eq{eigen_num} (color) versus analytic solutions \eq{eigen_ana} (dashed contours) at $\alpha=5$, $\kappa=12$, $q_{\rm Z}=0.3$, $u=10$, and $p_y=1$. The green curve shows $U$.  (b) Same mode structures in phase space: Wigner functions (color) versus isosurfaces of the  Hamiltonian \eq{Hamiltonian} (dashed contours). (c) Numerical (labeled ``N'') and analytic (labeled ``A'') growth rates versus $u$. The subscripts ``T'' and ``R'' stand for ``trapped'' and ``runaway''. The analytic growth rates are obtained by solving \Eq{eigen_ana} iteratively with $\omega=p_yU_0+p_y\kappa/\bar{p}_0^2-{\rm i}D_0$ as the initial guess. (d) Dimits shift in the Terry--Horton limit: simulation results of the modified Terry--Horton equation with $\delta_{\bd{p}}=\delta_0 p_y$ and $\hat{D}=1-0.01\nabla^2$ (green circles and red crosses) versus analytic results [\Eq{DS} with $p_y=1$ and $\varrho=0.05$]. Simulation details are the same as in \Ref{St-Onge17} (cf. Fig. 7 therein).}\label{TIstructure}
\end{figure}

To obtain  preliminary understanding of these eigenmodes, we first adopt a simple zonal profile:
\begin{equation}
U(x)=u\cos q_{\rm Z}x,\quad N(x)=0, \label{zonal_zeroN}
\end{equation}
in which case $\kappa_{\rm eff}=\kappa$. Then, numerical eigenmodes can be found from \Eq{eigen_num} assuming periodic boundary conditions. There are infinitely many eigenmodes, but most of them are small-scale and heavily damped. The two most unstable modes have the largest scales [\Fig{TIstructure}(a)]. These modes can be intuitively understood by examining their corresponding Wigner functions, $
W(x,p_x)\doteq\int {\rm d}s\,{\rm e}^{-{\rm i}p_xs}\psi(x+s/2)\psi^*(x-s/2)$, which loosely represents the distribution function of DW quanta, or ``driftons'' \cite{Smolyakov99,Zhu18b} (the asterisk denotes complex conjugate). In large-scale ZFs, driftons obey Hamilton's equations, where the Hamiltonian $H$ is obtained from $\hat{H}$ by replacing $(\hat{D},\hat{p}_x)$ with $(D_{\bd{p}}, p_x)$ and  taking the real part of the result \cite{Ruiz16,Zhu18b}:
\begin{equation}
H(x,p_x)=p_{y}[U+(\kappa+U''){\rm Re}\,\bar{p}^{-2}].\label{Hamiltonian}
\end{equation} 
Naturally, driftons tend to accumulate near phase-space equilibria of $H$, which are $x=x_n\doteq n\pi/q_{\rm Z}$, $p_x=0$ ($n=0,\pm 1, \pm 2, \dots$), so $W$ peaks near these locations (and overall, is aligned with isosurfaces of $H$), as seen in \Fig{TIstructure}(b). This explains eigenmode localization near  extrema of $U$. Maxima of $U$ (even $n$) correspond to phase-space islands encircled by ``trapped'' trajectories, and minima (odd $n$) correspond to saddle points passed by the ``runaway'' trajectories \cite{Zhu19,Ruiz16,Zhu18a,Zhu18b}. Hence, we call the modes localized near maxima and minima of $U$ as trapped and runaway modes, respectively. 

Let us consider a mode localized near some $x = x_n$ and shift the origin of $x$, so $U\approx U_0+\mc{C}x^2/2$, where $U_0=\pm u$ is the ZF velocity at the extremum and $\mc{C}\doteq U''(0)=\mp q_{\rm Z}^2u$ is the local ZF ``curvature''. Since the mode is localized in both $x$ and $p_x$,  one can Weyl-expand \cite{Dodin19}  $\hat{H}$ as $\hat{H}\approx\mc{H}+\lambda_p\,\hat{p}_x^2+\lambda_x\,x^2$, where
\begin{gather*}
\mc{H}=p_y [U_0 + (\kappa+\mc{C})\bar{p}_0^{-2}]-{\rm i}D_0,\quad \lambda_p=-p_y(\kappa+\mc{C})\bar{p}_0^{-4},\\
\lambda_x=\frac{p_{y}\mathcal{C}}{2}\left[ 1-\frac{p_{y}(\kappa+\mathcal{C})({\rm i}\alpha+p_{y}\kappa)}{\bar{p}_{0}^{4}({\rm i}\alpha+{\rm i}D_{0}+\omega-p_{y}U_{0})^{2}}\right],\\
\bar{p}_0^2=p_y^2+({\rm i}\alpha+p_y\kappa)/({\rm i}\alpha+{\rm i}D_0+\omega-p_yU_0),
\end{gather*}
and $D_0=D_{\bd{p}}(p_x=0)$. Then, \Eq{eigen_num} becomes similar to the  equation of a quantum harmonic oscillator
\begin{equation}
-\lambda_p\,\psi''+\lambda_x\,x^2\psi=(\omega-\mc{H})\psi,\label{Harmonic}
\end{equation}
except that its coefficients are complex. The solutions are \cite{Sakurai}
\begin{equation}
\psi_m={\rm e}^{-\frac{x^2}{2\lambda}}\mathsf{H}_m(x/\sqrt{\lambda}),\quad \omega_m=\mc{H}+(2m+1)\lambda_x\lambda,\label{eigen_ana}
\end{equation}
where $\mathsf{H}_m$ denotes Hermite polynomials, $\lambda\doteq\sqrt{\lambda_p/\lambda_x}$, and $m = 0, 1,2,...$. We choose the branch of the square root with ${\rm Re}\,\lambda>0$. The runaway and trapped modes shown in \Fig{TIstructure}(a) correspond to $m = 0$, and for comparison, we also plot our analytic solutions \eq{eigen_ana} in the same figure. Below, we only consider modes with $m=0$, which usually have the largest growth rates, and hence drop the subscript $m$. Also note that only the local ZF curvature $\mc{C}$ enters \Eq{Harmonic}, so the ZF does not need to be sinusoidal for \Eq{eigen_ana} to hold.

{\it TI as a modified primary instability.---} If $\gamma>0$, then the eigenmode \eq{eigen_ana} grows exponentially. This is the TI. Notice that in the absence of ZFs, $\omega$ reduces to $p_y\kappa/\bar{p}_0^2-{\rm i}D_0$, which is nothing but the primary-DW eigenfrequency \eq{LinearFre} at $p_x=0$. Therefore, the TI eigenmode can be understood as a standing DW eigenmode modified by the ZF, and from \Eq{eigen_ana}, its growth rate is
\begin{equation}
\gamma_{\rm TI}=\gamma_{\rm PI}+\Delta\gamma(\mc{C}),\quad\Delta\gamma=\gamma_{\rm PI}\,\mc{C}/\kappa+{\rm Im}(\lambda_x\lambda).\label{gammaTI}
\end{equation} 
Here, $\Delta\gamma=0$ at $\mathcal{C}=0$, when $\gamma_{\rm TI}$  reduces to $\gamma_{\rm PI}$. Also, $\Delta\gamma$ depends on the sign of $\mathcal{C}$, so the trapped and runaway modes have different growth rates. Predictions from the analytic formula \eq{gammaTI} are compared with numerical solutions of \Eq{eigen_num}  in \Fig{TIstructure}(c).  

These results show that mode localization is a key feature of the TI. This feature is missed in some previous studies \cite{St-Onge17,Kim02,Rath18} where the TI was derived from the interaction of just four Fourier harmonics. Also, our findings challenge the popular idea that the TI is a Kelvin--Helmholtz instability (KHI) \cite{Kobayashi12,Kim02,Numata07}.  
 Specifically, the TI studied here is caused by dissipation (\ie finite $\alpha$), just like the primary instability, while the KHI is caused by strong flow shear. Therefore,  unlike the KHI that develops when the ZF is too strong,  the TI develops when the ZF  is weak and becomes suppressed when the ZF is strong. The absence of KHI is due to the fact that, in the mHWE, $\tilde{\varphi}$ is assumed to have nonzero wavenumber $p_{\parallel}$ along $\vec{B}$ and thus electrons respond adiabatically ($\tilde{n} \approx \tilde{\varphi}$) at large $\alpha$. As shown in \Ref{Zhu18c}, for $\alpha \to \infty$, the KHI is suppressed at all $q_{\rm Z}^2 < 1$. Here, $\alpha$ is finite but still large ($\alpha = 5$), so similar conclusions apply. We found numerically (not shown) that for the parameters used in \Fig{TIstructure}(c), the KHI develops at $u\gtrsim 10^3$, which is too large to be relevant. Note that the assumption of nonzero $p_\parallel$ originates from the fact that the mHWE is intended to mimic turbulence in toroidal geometry with magnetic shear. But KHI can be more relevant in other geometries where $p_{\parallel}=0$ is possible and electron adiabatic response does not hold \cite{Hammett93}. For example, this is the case in Z pinches \cite{Ricci06}.

{\it TI and Dimits shift.---} The above TI theory  leads to a simple understanding of the Dimits shift, at least in the large-$\alpha$ limit. Assuming $\omega-p_yU\approx\kappa p_y/\bar{p}^2$ and small $D_0$, \Eq{pbarhat2} can be approximated  in this limit as
\begin{equation}
\hat{\bar{p}}^2=1+\hat{p}_x^2+p_y^2-{\rm i}\delta_{\bd{p}},\quad \delta_{\bd{p}}=\kappa p_y^2/[\alpha(1+p_y^{2})].\label{THmodel}
\end{equation}
This corresponds to the Terry--Horton  model \cite{Terry82,St-Onge17}, which is a one-field model that evolves $\varphi$ but not $n$ (\ie $N=0$) and assumes a constant phase shift $\delta_{\bd{p}}$ between $\tilde{\varphi}$ and $\tilde{n}$. (Different forms of $\delta_{\bd{p}}$  can also be chosen to model different primary instabilities \cite{St-Onge17}.) Then, the Hamiltonian \eq{eigen_num} no longer depends on $\omega$, so the growth rate is found explicitly:
\begin{equation}
\gamma_{\rm TI}={\rm Im}\left[\frac{p_y(\kappa+\mc{C})}{1+p_y^2-{\rm i}\delta_{\bd{p}}}\left(1-\sqrt{\frac{\mc{C}}{2(\kappa+\mc{C})}}\right)\right]-D_0.\label{gammaTI_TH}
\end{equation}
Here, we consider the runaway mode ($\mc{C}>0$), because it is usually more unstable than the trapped mode in this model. Equation~(\ref{gammaTI_TH}) allows calculation of the TI threshold, \ie the value of $\kappa$ at which $\gamma_{\rm TI} = 0$. We denote this value as $\kappa_{\rm c}$. One finds from \Eq{gammaTI_TH} that 
$\kappa_{\rm c}$ differs from the linear threshold by $\Delta_{\rm DS}$:
\begin{equation}
\Delta_{\rm DS}=\kappa_{\rm c}-\kappa_{\rm lin},\quad\kappa_{\rm c}=\frac{D_0}{p_y}\,\frac{(1+p_y^2)^2+\delta_{\bd{p}}^2}{\delta_{\bd{p}}-(1+p_y^2)\sqrt{\varrho/2}},\label{DS}
\end{equation}
where $\kappa_{\rm lin}\doteq\kappa_{\rm c}|_{\varrho=0}$ is the linear threshold of the primary instability  and $\varrho \doteq \mathcal{C}/\kappa$. (To simplify the formula for $\kappa_{\rm c}$, we assumed the typical regime $\varrho\ll 1$.) 

A ZF cannot suppress the TI at $\kappa>\kappa_{\rm c}$, so $\Delta_{\rm DS}$ is the Dimits shift.   For ZFs generated self-consistently,  $\varrho$ is roughly constant (see below) and can be estimated numerically. Then, $\Delta_{\rm DS}$ is found by minimizing \Eq{DS} over $p_y$. Following \Ref{St-Onge17}, we adopt $\delta_{\bd{p}}=\delta_0 p_y$ and $\hat{D}=1-0.01\nabla^2$; then, $\Delta_{\rm DS}$ is minimized at $p_y\approx 1$ and its corresponding value is in good agreement with direct simulations of the modified Terry--Horton system [\Fig{TIstructure}(d)]. The simulation details are the same as in \Ref{St-Onge17}, where similar numerical results are compared with a different theory. Unlike in \Ref{St-Onge17}, we derive $\Delta_{\rm DS}$ explicitly and do not reduce the problem to the interaction of just four Fourier harmonics.  Furthermore, our $\gamma_{\rm TI}$ is determined by the \textit{local} ZF curvature, so $\Delta_{\rm DS}$ is insensitive to the specific shape of the ZF.  

In summary, in the Terry--Horton limit, the Dimits shift is caused by the difference between $\gamma_{\rm TI}(\mathcal{C})$ and $\gamma_{\rm PI}$, and the Dimits regime ends when the ZF becomes \textit{too weak} to suppress the TI. This is different from the previous study \cite{Rogers00} of the ITG system, which found that the TI becomes unstable when the ZF becomes too strong. The difference is due to the fact that in \Ref{Rogers00}, the perturbations of the ion perpendicular temperature cause an additional destabilizing effect. However, the stabilizing effect of the ZF curvature remains the same \cite{Supplement}.

{\it TI's roles in nonlinear dynamics.---} At smaller $\alpha$,  $N$ is nonzero \cite{foot} and can affect $\gamma_{\rm TI}$ and $\Delta_{\rm DS}$. This makes the mHWE model more complicated than its Terry--Horton limit considered above. Nevertheless, some interesting phenomena seen in mHWE simulations can be explained in the context of our TI theory.

\begin{figure}
\includegraphics[width=1\columnwidth]{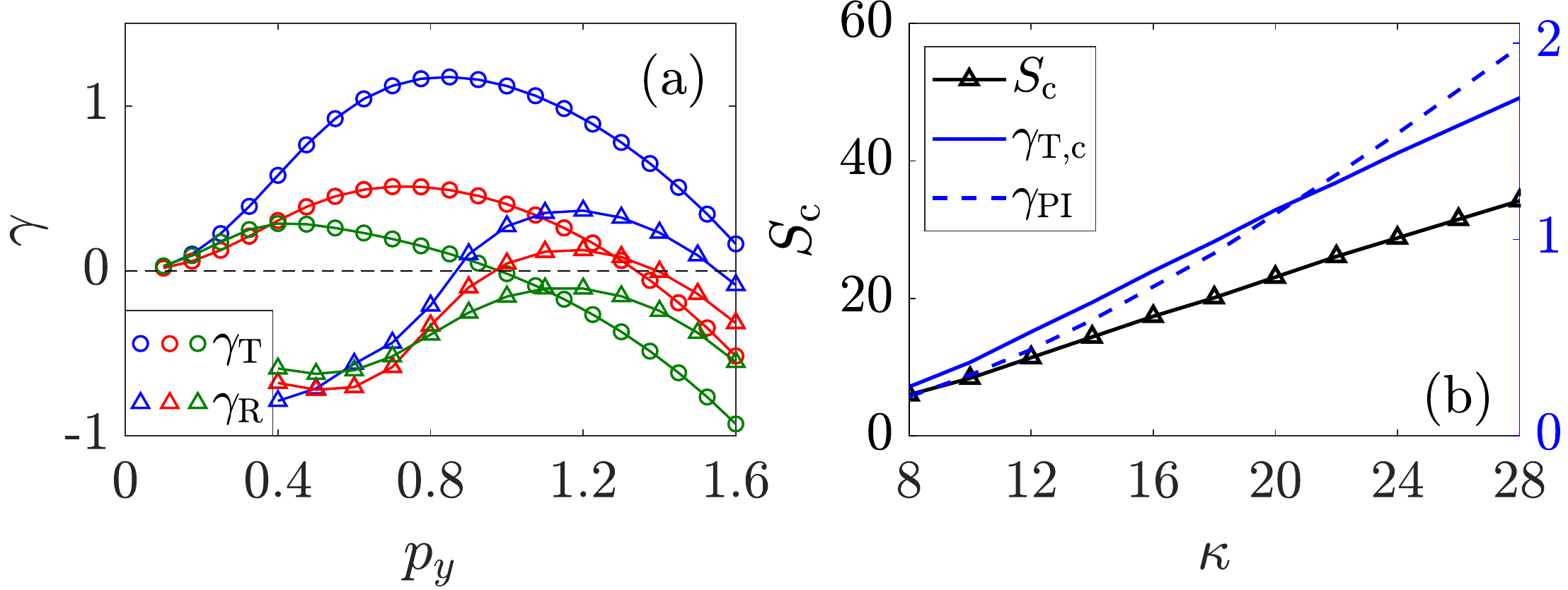}
\caption{(a) The growth rates of the trapped ($\gamma_{\rm T}$, circles) and runaway ($\gamma_{\rm R}$, triangles) modes found numerically for the zonal profile \eq{zonal_nonzeroN} at $q_{\rm Z}=0.3$ and $u=20$. Different colors indicate different values of $N_0$ for trapped (runaway) modes: blue, $N_0=0$ ($N_0=0$); red, $N_0=5.4$ ($N_0=1.8$); green, $N_0=9$ ($N_0=3.6$). The peak of $\gamma_{\rm T}$  shifts to smaller $p_y$, but the peak of $\gamma_{\rm R}$  remains at $p_y\approx 1.2$. (b) The critical shear $S_{\rm c}=q_{\rm Z}u_{\rm c}$ versus $\kappa$ at $\alpha=5$, $q_{\rm Z}=0.3$, and $p_y=0.4$. We also plot $\gamma_{\rm PI}$,  and  $\gamma_{\rm T}$ at $u=u_c$ (denoted $\gamma_{\rm T,c}$).}\label{ShearLevel}
\end{figure}

To simplify the mode structure, we assume large hyperviscosity, so small-scale variations of $U$ and $N$ are washed out. (Using normal viscosity leads to similar results, as  checked numerically.) In this case, trapped and runaway modes are still observed yet exhibit different $p_y$ [\Fig{SnapShot}(a)]. This can be explained by allowing for nonzero $N$. For example, we consider
\begin{equation}
U(x)=u\cos q_{\rm Z}x,\quad N(x)=N_0\sin 2q_{\rm Z}x \label{zonal_nonzeroN}
\end{equation}
based on numerical observations [\Fig{SnapShot}(a)]. The corresponding growth rates $\gamma$ of the trapped and runaway modes are found numerically. It is seen in  \Fig{ShearLevel}(a) that these growth rates decrease as $N_0$ increases (\ie as $\kappa_{\rm eff}$ decreases at extrema of $U$). For the runaway mode, the peak of $\gamma$ remains at $p_y\approx 1.2$; but for the trapped mode,  the peak shifts to smaller $p_y$. This agrees with the self-consistent simulations showing that the trapped mode has smaller $p_y$ [\Fig{SnapShot}(a)].

\begin{figure}
\includegraphics[width=1\columnwidth]{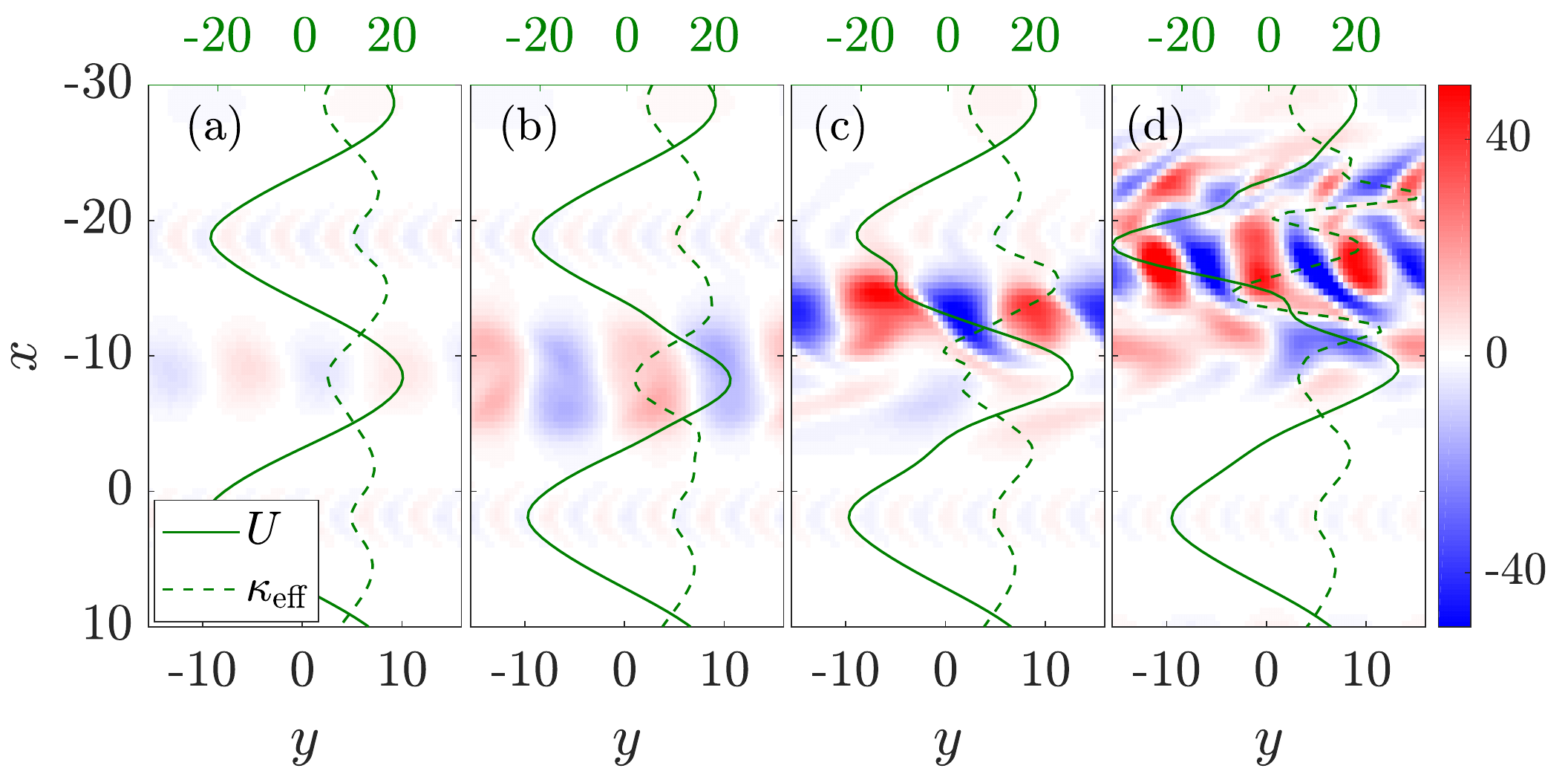}\caption{Snapshots of a propagating structure corresponding to: (a) $t=756$, (b) $t=763$, (c) $t=770$, and (d) $t=777$ of \Fig{SnapShot}. The axes are rotated  by $90^{\circ}$ compared with \Fig{SnapShot}(a). This structure originates as a trapped mode ($p_y=0.4$) at $x\approx -10$ and propagates to the neighboring minimum of $U$ at $x\approx -20$. Both the minimum and the maximum of $U$ are amplified by this structure, resulting in an increase of $E_U$.}\label{PropStructure}
\end{figure}

Since ZFs are subject to viscous damping, turbulence cannot be suppressed indefinitely, and PP oscillations occur. Compared to those in zero-dimensional models \cite{Diamond05}, the PP oscillations are more intricate and can  be of two types: type-I corresponds to the exchange between $E_{\rm DW}$ and $E_{N}$, while type-II corresponds to the exchange between $E_{\rm DW}$ and $E_{U}$ [\Fig{SnapShot}(b)]. Type-I PP oscillations  occur more frequently  because $E_N$ decays faster than $E_U$ ($N$ is more prone to viscous damping due to larger wavenumber), and the corresponding bursts of $E_{\rm DW}$ saturate quickly due to the decrease of $\kappa_{\rm eff}$. The latter increases $E_{N}$  but has little effects on $E_U$, so the ZF amplitude $u$ decreases approximately monotonically. However, when $u$ becomes small enough, a type-II PP oscillation  occurs. Then, a trapped mode does not saturate but develops into an avalanche-like nonlinear structure propagating radially (\Fig{PropStructure}). This propagation triggers a turbulence burst ending with a rapid increase of the ZF amplitude; hence, $E_U$ increases too. After such a burst, turbulence becomes suppressed again, and the Dimits regime is reinstated until the next type-II PP oscillation. (Similar propagating structures have also been found in a  reduced model of ITG turbulence \cite{Ivanov19} and separately, in simulations of DW turbulence in a linear device \cite{Lang19}.)

Note that the propagating structures observed here are different from the DW--ZF solitons generated in the large-$\alpha$ limit \cite{Zhou19a}. They may be related to those seen in simulations of subcritical turbulence \cite{Zhou19b,vanWykl16,McMillan18}. In our case,  for a given $q_{\rm Z}$, whether the trapped mode can develop into a propagating structure depends on the critical ZF amplitude $u_{\rm c}$, or the critical shear $S_{\rm c}\doteq q_{\rm Z}u_{\rm c}$. The critical shear in turn determines the characteristic ZF amplitude, \ie $u\approx u_{\rm c}$, because ZFs decay at $u > u_{\rm c}$ and are amplified by propagating structures at $u < u_{\rm c}$.

We numerically identify the critical shear  by considering the initial zonal profile (\ref{zonal_zeroN}) and applying a small perturbation $\tilde{w}\propto{\rm e}^{{\rm i}p_y y}$. By varying $u$, we find the critical value $u=u_{\rm c}$ and the corresponding critical shear $S_{\rm c}=q_{\rm Z}u_{\rm c}$ above which the structure ceases to propagate. The results are shown in \Fig{ShearLevel}(b), where we also plot $\gamma_{\rm PI}$ and $\gamma_{\rm T,c}$, the latter being the trapped-mode growth rate at $u=u_{\rm c}$. It is seen that $\gamma_{\rm PI}$ is not a linear function of $\kappa$ (rather, $\gamma \sim \kappa^{1.5}$), but both $S_{\rm c}$ and $\gamma_{\rm T,c}$ increase linearly with $\kappa$. This justifies our earlier assumption that $\varrho$ is constant in \Eq{DS}.

The effects of $S$ (typical $U'$) and $\mc{C}$ (typical $U''$) together determine the characteristic ZF wavenumber $q_{\rm Z}$ in the Dimits regime, specifically, as follows.  First, $q_{\rm Z}$ cannot be too large, because large $q_{\rm Z}$ corresponds to small ZF amplitude $u\sim|\mc{C}|/q_{\rm Z}^2$ assuming $|\mc{C}|$ is bounded by the TI threshold (otherwise, the TI is stable and the ZF decays), and ZFs with large $q_{\rm Z}$ and small $u$  tend to merge \cite{Zhu18b,Zhu19,Parker14}. Next, $q_{\rm Z}$ cannot be too small either, because small $q_{\rm Z}$ corresponds to small $|\mc{C}|\sim q_{\rm Z}S_{\rm c}$ (assuming $S\sim S_{\rm c}$), unleashing the primary instability; then, the secondary instability would develop like in homogeneous turbulence and the ZF with larger $q_{\rm Z}$ would emerge. 

As $\kappa$ increases, the TI becomes more active in producing propagating structures. When such structures are produced almost continuously, the Dimits regime ends, i.e., plasma becomes turbulent. As found numerically (not shown), for $\alpha = 5$, that occurs at $\kappa\gtrsim 20$. This threshold does not depend significantly on the simulation-domain size as long as the latter is large enough compared to the typical scales of ZFs and DWs.

{\it Discussion.--- } Although we adopt the mHWE as our base model, our general approach to the TI is applicable more broadly. Any other model also leads to \Eq{eigen_num}, except with a different $\hat{H}$. Since the TI localization is a feature common in many models \cite{Rogers00,Rogers05,Kobayashi12,Kim18a,Kim19,Ivanov19}, one can Weyl-expand the corresponding $\hat{H}$ like we did above and again arrive at \Eq{eigen_ana} (with different coefficients); hence, the qualitative physics remains the same. In Supplemental Material \cite{Supplement}, we show how this approach helps reproduce the key features of the TI in ITG turbulence within the model studied in \Ref{Rogers00}. We also show there that gyrokinetic simulations of ITG turbulence (using the code \textsc{GS2} \cite{GS2}) exhibit similar localized modes. Although these supplemental findings are not central to our work, they suggest interesting directions for future research.

\begin{acknowledgments}
This work was supported by the US DOE through Contract No. DE-AC02-09CH11466. This work made use of computational support by CoSeC, the Computational Science Centre for Research Communities, through CPP Plasma (EP/M022463/1) and HEC Plasma (EP/R029148/1).
\end{acknowledgments}

\end{document}